\documentstyle[epsfig,fleqn,rotating]{basi}
\begin{document}
\title[] 
      {Aerosol contents at an altitude of ~2 km in central Himalayas}
\author[]{Ram Sagar,$^{1}$\thanks{e-mail: sagar@upso.ernet.in} B. Kumar,$^{1}$ P. Pant,$^{1}$ U.C. Dumka,$^{1}$
K.K. Moorthy$^{2}$ 
\newauthor
and R. Sridharan$^{2}$  \\ \\
$^{1}$State Observatory, Manora Peak, Nainital 263 129, India\\
$^{2}$Space Physics Laboratory, Vikram Sarabhai Space Centre, Thiruvananthapuram 695 022, India}
\maketitle
\label{firstpage}

\section{Introduction}
\hskip 10pt Aerosols, both natural and anthropogenic, play an important role in the atmospheric science, by imparting 
radiative forcing and 
perturbing the radiative balance of the Earth atmosphere 
system as well as by degrading the environment. To understand the effect of aerosols on our geo/biosphere system, it 
is essential to characterize their physical and chemical properties regionally because of the regional nature of their 
properties and the short lifetime (Moorthy et al. 1999, 1989, Satheesh et al. 2001). As most of the aerosol sources 
are of terrestrial origin the 
variability of their properties will be very large close to the surface. At higher altitudes, above the mixing region, and 
in the free troposphere, the aerosol characteristics have a more synoptic perspective; would be indicative of the 
background level and are useful to understand long-term impacts. Such systematic measurements of aerosols at high altitudes 
are practically non-existing in India. Realising the potential and need for such studies, an activity has been initiated 
at Manora Peak, Nainital in the Shivalik Hills of Central Himalayas at an altitude of $\sim$2 km. 
The present paper provides the preliminary results of these aerosol measurements. These daytime measurements are compared 
with the existing earlier nighttime measurements. 

\section{Observational data and analysis}
The daytime observations for aerosol optical depths were made for 68 clear-sky days during January 2002 to June 2002, with
Multi-Wavelength Solar Radiometer (MWR) designed and developed by the Space Physics Laboratory (SPL), Thiruvananthapuram.
The MWR was installed at Manora Peak, Nainital (1965 m msl). The instrument provides aerosol spectral optical depth at ten 
narrow (FWHM of 6 to 10 nm) spectral bands centered at 380, 400, 450, 500, 600, 650, 750, 850, 935, and 1025 nm, by making
continuous spectral extinction measurements of directly transmitted solar radiation (Moorthy et al. 1999). 
The nighttime extinction measurements were carried out regularly at the observatory using broad band (FWHM $\sim$ 60 nm) 
filters with effective wavelengths around 360, 450 and 550 nm using stellar sources. These mesurements are available for about 800 
nights between 1964 to 1988 (Kumar et al. 2000).    

\begin{figure}[t]
\centerline{\psfig{file=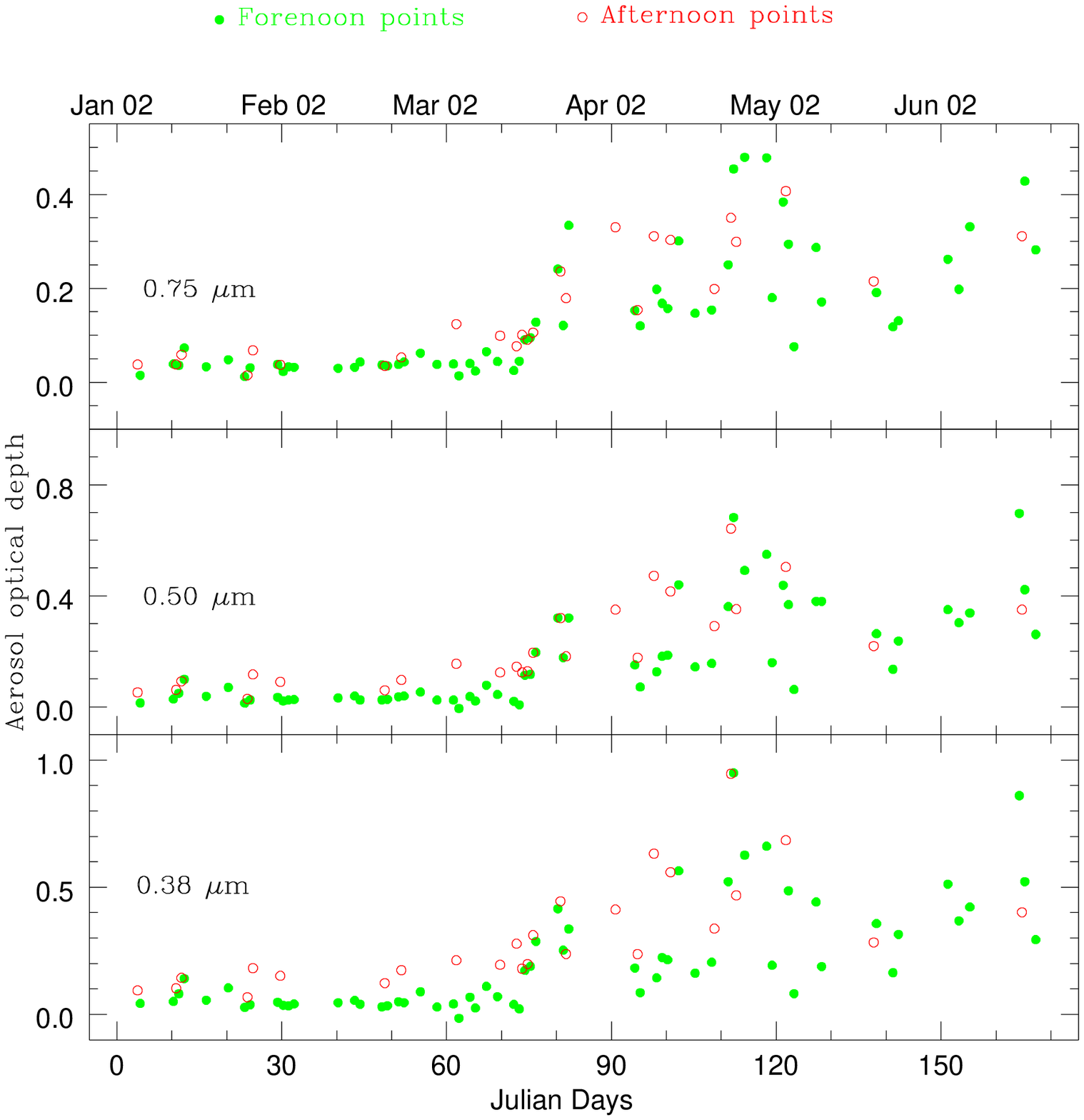, height=10cm}}
{{\bf Figure 1.} Temporal variation of the daytime aerosol optical depth.} 
\end{figure}

Each set of MWR observation comprises time measured with seconds accuracy and the output voltage V$_{\lambda}$. The data
are recorded from morning to evening and optical depths are estimated following the Langley technique (Moorthy et al. 1999).
A preliminary analysis of the aerosol measurements indicated the occurence of different least squares fit to the data on 
several days, with different slopes for the forenoon (FN) and afternoon (AN) parts of the same day implying different 
optical depths. As such, these data were analysed separately and in total we have 90 data sets spanning over the 68 
observation days spread over nearly 6 months.    

\section{Results and discussions}
The temporal variation of the daytime aerosol optical depths estimated using the MWR are shown 
in Fig. 1 as a function of the Julian days. Except a few days in April, FN AODs are lower by 0.02 to 0.2 than the 
corresponding AN values. With a view to examine the temporal variation with the earlier nighttime values, the 
archived nighttime aerosol measurements (broad bands) at 0.36 $\mu$m and 0.44 $\mu$m, were used. The individual values were 
grouped into fortnightly ensembles and averaged separately for the daytime data (at 380 nm and 450 nm) and nighttime 
data (at 360 nm and 440 nm). The temporal variations are shown in Fig. 2; on the top panel of which the 
AOD at 750 nm from MWR are also given. The wavelength dependence of aerosol optical depths in different months for 
the MWR data is plotted in Fig. 3. 

\begin{figure}[t]
\centerline{\psfig{file=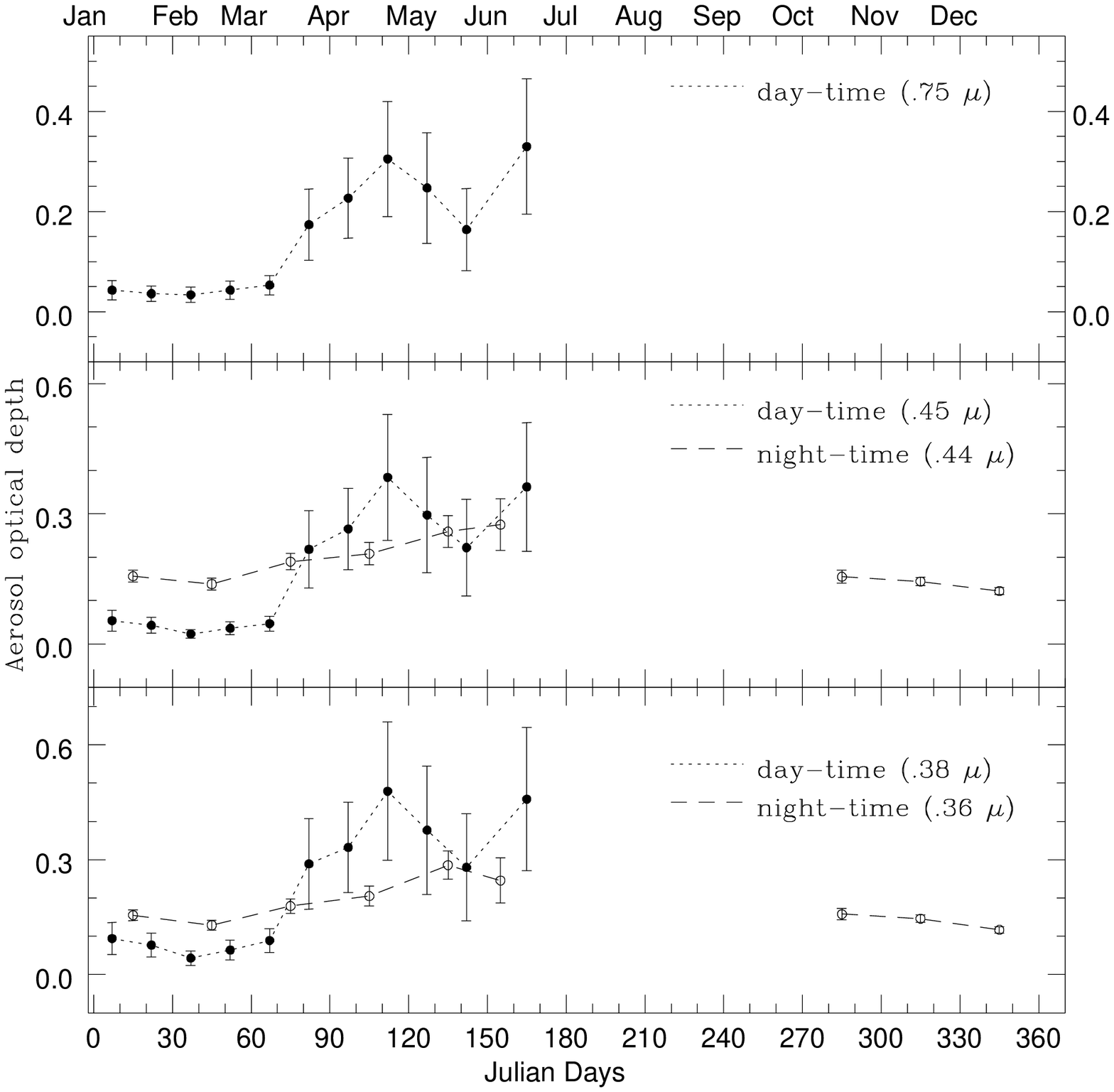,height=10cm}}
{{\bf Figure 2.} Temporal variation of mean aerosol optical depths during the year at Nainital.} 
\end{figure}

The primary features of the aerosol optical depths (Fig. 1) are (i) low values in winter; (ii) a rather spontaneous increase
from spring and (iii) reaching a peak by summer; when the AOD values are nearly 4 times the winter values. 
The increase is more spontaneous at the longer wavelengths ($0.75\mu$m), while at the shorter wavelengths 
it is more gradual. This is attributed to increased convective activity in summer taking place in the lower plains due to 
increased solar heating and the resulting aerosols are lifted to higher altitudes by turbulent exchange. As the turbulence 
increases towards summer, even larger particles are lifted up causing a spontaneous increase in the AOD at the long 
wavelength also. 
Another feature, the FN AOD being lower than the AN values, arises due to the environmental difference; in the FN, the solar ray 
path is primarily through the cleaner mountain ranges in the east whereas in the evening the dusty and inhabitated valley 
intercepts the ray path. Both the daytime and nighttime AOD's show similar trends; except that the nighttime AOD's tend to be 
higher than the daytime counter parts during winter and vice versa in summer. This aspect and its relation to the micrometerology 
as well as the effect of the bandwidth need further investigation.

\begin{figure}[t]
\centerline{\psfig{file=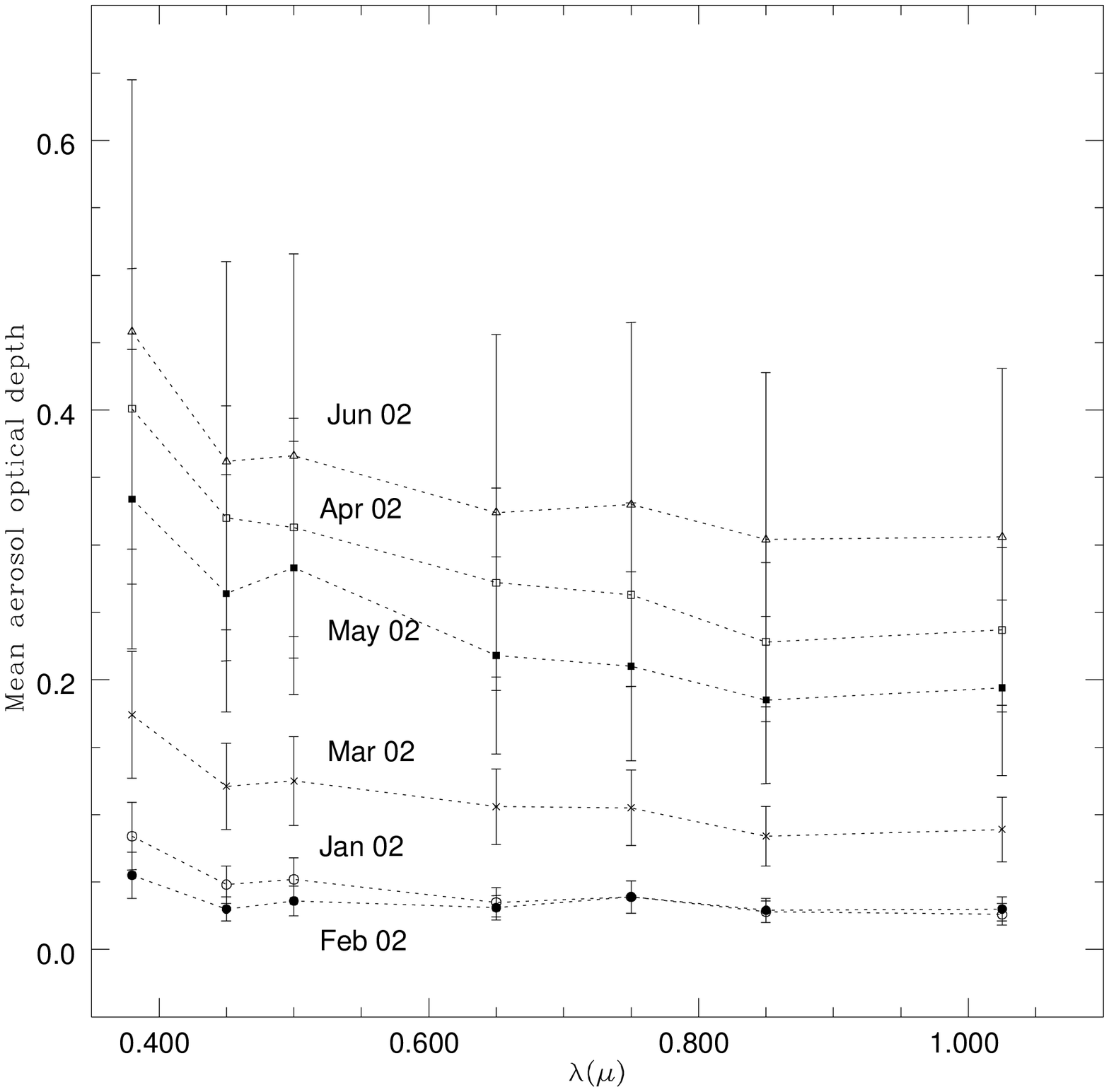,height=10cm}}
{{\bf Figure 3.} Wavelength dependence of the aerosol optical depth.} 
\end{figure}

\section*{Acknowledgements}
The authors thank the technical staff of the State Observatory, Nainital for providing valuable help during observations.

\label{lastpage}
\end{document}